\documentclass[preprint2]{aastex}
\usepackage{verbatim}
\usepackage{gensymb}

\newcommand{\noprint}[1]{}

\newcommand{\solarmass}{M_{\sun}}

\begin{document}

\title{Turbulence and Star Formation in a Sample of Spiral Galaxies}
\author{Erin Maier}
\email{erin-maier@uiowa.edu}
\affil{Department of Physics and Astronomy, Northern Arizona University}
\affil{527 S Beaver Street, Flagstaff, AZ 86011}
\affil{Department of Physics and Astronomy, The University of Iowa}
\affil{30 N Dubuque Street, Iowa City, IA 52242}
\author{Li-Hsin Chien}
\email{Lisa.Chien@nau.edu}
\affil{Department of Physics and Astronomy, Northern Arizona University}
\affil{527 S Beaver Street, Flagstaff, AZ 86011}
\author{Deidre A. Hunter}
\email{dah@lowell.edu}
\affil{Lowell Observatory}
\affil{1400 W Mars Hill Road, Flagstaff, AZ 86001}

\shorttitle{Turbulence and Star Formation in Spiral Galaxies}
\shortauthors{Maier, Chien, and Hunter 2016}

\begin{abstract}
We investigate turbulent gas motions in spiral galaxies and their importance to star formation in far outer disks, where the column density is typically far below the critical value for spontaneous gravitational collapse. Following the methods of \citet{Burk} on the Small Magellanic Cloud, we use the third and fourth statistical moments, as indicators of structures caused by turbulence, to examine the neutral hydrogen (\ion{H}{1}) column density of a sample of spiral galaxies selected from The \ion{H}{1} Nearby Galaxy Survey \citep[THINGS,][]{Walter}. We apply the statistical moments in three different methods--- the galaxy as a whole, divided into a function of radii and then into grids. We create individual grid maps of kurtosis for each galaxy. To investigate the relation between these moments and star formation, we compare these maps with their far-ultraviolet images taken by the {\it Galaxy Evolution Explorer} ($GALEX$) satellite. We find that the moments are largely uniform across the galaxies, in which the variation does not appear to trace any star forming regions. This may, however, be due to the spatial resolution of our analysis, which could potentially limit the scale of turbulent motions that we are sensitive to greater than $\sim700$~pc. 
From comparison between the moments themselves, we find that the gas motions in our sampled galaxies are largely supersonic. This analysis also shows that \citet{Burk}'s methods may be applied not just to dwarf galaxies but also to normal spiral galaxies. 
\end{abstract}


\section{INTRODUCTION \label{intro}}

Turbulence in interstellar medium (ISM) has been shown as the dominant source of structure and dynamics of the various gas phases \citep[e.g.~][]{larson81,dl90,burkert06}. When temperature drops below $10^4$~K in the interstellar medium, the kinetic turbulent pressure becomes important as it often exceeds the thermal pressure. The sonic Mach number, $\mathcal{M}_{s}$, can be an indicator of the medium in which turbulent motions are taking place. A Mach number of $4$ to $5$ has been commonly associated with the cold neutral medium where molecular clouds form, and a Mach number of $\sim1.4$ to $2.4$ has been associated with the warm ionized medium in the Milky Way Galaxy. Turbulence is driven on different length scales, from star-forming molecular clouds to galactic spiral arms, by various energetic sources such as stellar feedback, high \ion{H}{1} velocity dispersion in parts of galactic disks, galactic rotation, and the magneto-rotational instability caused by the coupling of galactic shear with magnetic field \citep[e.g.~][]{elm02,henne12}. However, \citet{Stilpc} found that no theorized trigger of turbulence in neutral hydrogen gas (\ion{H}{1}) is powerful enough to drive the observed levels on its own.

Many studies suggest that turbulence compression is one of the triggers of star formation in clouds. The instabilities due to localized cloud collapses induce turbulence that compresses the gas further, which can cover scales below the Jeans length, and cascade downward quickly to produce very small structure in the cool ISM \citep{elm02}. Such theoretical models of gas motions and compressions in a turbulent fluid shows that when power spectrum analysis is done on the gas emissions, we can expect a power law relation of the spatial frequencies \citep{goldman,lp00}. Therefore, power spectrum analysis of emission maps is often used as an observational diagnostic for the dynamics and stability of ISM \citep{es04}, both at galaxy scales and cloud scales. Based on these observational analyses \citep[e.g.~][and references within]{elm01,block,combes12,henne12}, the power spectra can be approximated as two power laws, a shallow one on large scales ($>100$~pc) and a steeper one on small scales, with the break between the two corresponding to the galaxy's line-of-sight disk thickness. On large scales, turbulent processes and gas motions are approximately two dimensional, driven by density waves in the disk, while on small scales, are three dimensional, controlled by star formation and feedback (possibly through expanding \ion{H}{2} regions, although this is not well understood). On the other hand, gas turbulence can also disrupt gaseous structures faster than they gravitationally collapse, which could potentially quench the star formation activity in galaxies \cite[e.g.~][]{dekel,martig,murray,os11}.

In low density regions in galaxies, such as the outer skirts in galactic disks and dwarf irregular galaxies \citep[e.g.~][]{Elme, Kral}, turbulence may play an even more crucial role in causing star formation. However, the mechanism of turbulence in the inner, higher density, and outer, lower density regions of spiral galaxies could be different \citep{Pion, Tamb}, and such differences could also contribute to differences in the presence and efficiency of star formation \citep{Bourn, Ren}. In this study we seek to understand the role of turbulence in star formation in normal spiral galaxies overall and in their outer disks, using a new analysis based on statistical moments.

Several statistical methods have proven useful in analyzing turbulent structure in ISM \citep[e.g.~][]{Kowal, Laz, bfkl09, Burk}. In particular, \citet{Burk} used a combination of observational data and a database of simulations to show that the statistical moments---variance, skewness and kurtosis, when applied to the \ion{H}{1} column density images, can be used to characterize the turbulent structure in the Small Magellanic Cloud. Specifically, these statistical moments have a dependence on the sonic Mach number, $\mathcal{M}_{s}$. 
Variance has a linear dependence over a broad range of $\mathcal{M}_{s}$, while both skewness and kurtosis have a rather flat but still increasing dependence on $\mathcal{M}_{s}$ for supersonic models. These dependences can be explained in that as $\mathcal{M}_{s}$ increases, so does the Gaussian asymmetry of the column density (and density) distributions due to higher gas compression via shocks, resulting in the increase of variance, skewness, and kurtosis. However, \citet{Burk} suggested that, in order to make a direct comparison between simulations and observations, the higher order moments, skewness and kurtosis, are more appropriate statistics to use since they are not affected by the scaling of the data set. Moreover, when pixel-to-pixel comparison between skewness and kurtosis is performed, \citet{Burk} found that the supersonic model shows a good correlation between kurtosis and skewness, while the subsonic model does not (see Section~\ref{HOS}).

In this study, we apply the technique from \citet{Burk} to a sample of ten galaxies chosen from The \ion{H}{1} Nearby Galaxy Survey \citep[THINGS\footnote{See http://www.mpia.de/THINGS/Overview.html.},][]{Walter} to test whether it is applicable to normal spiral galaxies. We first apply the third and fourth order statistical moments, skewness and kurtosis, to \ion{H}{1} column density images of our galaxies in order to search for signs of turbulence. We then compare these high-order statistics (HOS) to the photometric variations in each galaxy's far-ultraviolet (FUV) image taken by the {\it Galaxy Evolution Explorer} ($GALEX$) satellite. As FUV emission is a direct indicator of young stars, a comparison between the FUV emission to the variations in the HOS of the \ion{H}{1} gas may reveal a relationship between star formation and turbulence. 

This paper is organized as the follows: \S\ref{data} describes our data set, \S\ref{analysis} presents our statistical methods and the three different ways we applied these methods to the galaxies, and \S\ref{discuss} summarizes our results and discusses some possible further analysis.

\section{DATA \label{data}}
\subsection{THINGS \ion{H}{1} Column Density Images}


\begin{deluxetable}{llccccccccc}
\tablecaption{Properties of Sampled Galaxies\label{galprop}}
\tablewidth{0pt}
\tabletypesize{\footnotesize}
\tablehead{
\colhead{} & \colhead{R.A. (J2000)} & \colhead{Dec. (J2000)} & \colhead{$D$} & \colhead{$R_{25}$} & \colhead{Incl.} & \colhead{P.A.} & \colhead{} & \colhead{E(B$-$V)} & \colhead{log(SFR)} \\
\colhead{Name} & \colhead{(hh mm ss.s)} & \colhead{(dd mm ss)} & \colhead{(Mpc)} & \colhead{(kpc)} & \colhead{(deg)} & \colhead{(deg)} & \colhead{$b/a$} & \colhead{(Mag)} & \colhead{($\solarmass$\,yr$^{-1}$\,kpc$^{-2}$)} \\
\colhead{(1)} & \colhead{(2)} & \colhead{(3)} & \colhead{(4)} & \colhead{(5)} & \colhead{(6)} & \colhead{(7)} & \colhead{(8)} & \colhead{(9)} & \colhead{(10)}
}

\startdata

NGC 925 & 02 27 16.5 & 33 34 44 & 9.2 & 14.3 & 66 & 287 & 0.45 & 0.067 & $-$2.77\\
NGC 2403 & 07 26 51.1 & 65 36 03 & 3.2 & 7.38 & 63 & 124 & 0.49 & 0.035 & $-$2.30\\
NGC 2976 & 09 47 15.3 & 67 55 00 & 3.6 & 3.80  & 65 & 335 & 0.46 & 0.063 & $-$2.66\\
NGC 3031  & 09 55 33.1 & 69 03 55 & 3.6 & 11.2 & 59 & 330 & 0.33 & 0.071 & $-$2.57\\ 
NGC 4736 & 12 50 53.0 & 41 07 13 & 4.7 & 5.19 & 41 & 296 & 0.77 & 0.015 & $-$2.31\\
NGC 5194 & 13 29 52.7 & 47 11 43 & 8.0 & 9.04 & 42 & 172 & 0.76 & 0.031 & $-$1.63\\
NGC 5236 & 13 37 00.9 & $-$29 51 57 & 4.5 & 10.1 & 24 & 225 & 0.92 & 0.059 & $-$2.11\\
NGC 6946  & 20 34 52.2 & 60 09 14 & 5.9 & 9.86 & 33 & 243 & 0.85 & 0.303 & $-$1.81\\
NGC 7793 & 23 57 49.7 & $-$32 35 28 & 3.9 & 5.94 & 50 & 290 & 0.66 & 0.017 & $-$2.34\\
IC 2574 & 10 28 27.7 & 68 24 59 & 4.0 & 7.50 & 53 & 56 & 0.62 & 0.032 &  $-$3.17\\
\enddata

\tablecomments{Column~4: distance taken from NED; Column~5: optical size calculated at distance $D$ based on angular $R_{25}$ taken from LEDA; Column~6 \& 7: inclination and position angle taken from LEDA; Column~8: semi-minor to semi-major axis ratio, calculated from inclination assuming an intrinsic axis ratio of $0.2$ for normal spiral galaxies; Column~9: foreground galactic extinction taken from NED for the recalibration of the \citet{schlegel98} values by \citet{schlafly11}; Column~10: normalized star formation rate per area of the galaxy, SFR adopted from \citet{Walter} and area estimated from $R_{25}$.}

\end{deluxetable}

Table~\ref{galprop} summarizes the properties of the selected ten galaxies, which cover a distance of $3.2$ to $9.2$~Mpc and types of Sab to Sd spirals. These galaxies were selected from the THINGS sample because they have beam sizes that are less than or equal to $\sim200$~pc. We utilize the {\sc robust}-weighted \ion{H}{1} column density (moment $0$) images from THINGS, a survey of nearby galaxies obtained at the Very Large Array (VLA) of the National Radio Astronomy Observatory\footnote{The National Radio Astronomy Observatory is a facility of the National Science Foundation operated under cooperative agreement by Associated Universities, Inc.} \citep{Walter}. These images have beam sizes of $4.8\arcsec$ to $10.4\arcsec$. However, in order to analyze the galaxies at the same intrinsic spatial scale, we smoothed the images to a resolution of $200$~pc for all galaxies. Note that we found the pixel scale of the acquired NGC~5236 image is 1.5$\arcsec$ per pixel and the original image size is $1024 \times 1024$ pixels, different from what is recorded in \citet{Walter}. For the NGC~3031 image, several bright continuum sources and noisy regions at the edges of the image were removed in order to reduce errors in the following statistical analysis.

\subsection{$GALEX$ Far-Ultraviolet Images}

We also utilize FUV ($\lambda_{eff} = 1516$~\AA) images from the Nearby Galaxy Survey \citep{NGS} obtained with the $GALEX$ satellite \citep{GALEX}\footnote{NGC~6949 is not in the NGS survey.}. Each FUV image was first trimmed to a size of $2000 \times 2000$ pixels where the pixel scale is 1.5$\arcsec$ per pixel, while NGC~925 and NGC~7793 have smaller image sizes, $557 \times 557$ and $629 \times 629$ pixels respectively, and were not trimmed. In order to perform surface photometry, we first removed any unassociated galaxies and foreground stars from the image, so that the sky background could be fit and subtracted. The foreground extinction was corrected using E(B$-$V) values listed in Table~\ref{galprop}, however we did not correct for reddening internal to the galaxies. This would serve to brighten the FUV photometry and introduce a radial trend since reddening should decrease with distance from the center of the galaxy. However, our analysis of comparing FUV variations with the \ion{H}{1} HOS would not be affected, since features in the FUV profiles would remain. For NGC~3031, a nearby dwarf galaxy was masked out in order to show the surface brightness of NGC~3031 alone. 

\begin{figure}[t!]
\epsscale{0.8}
\centering
\plotone{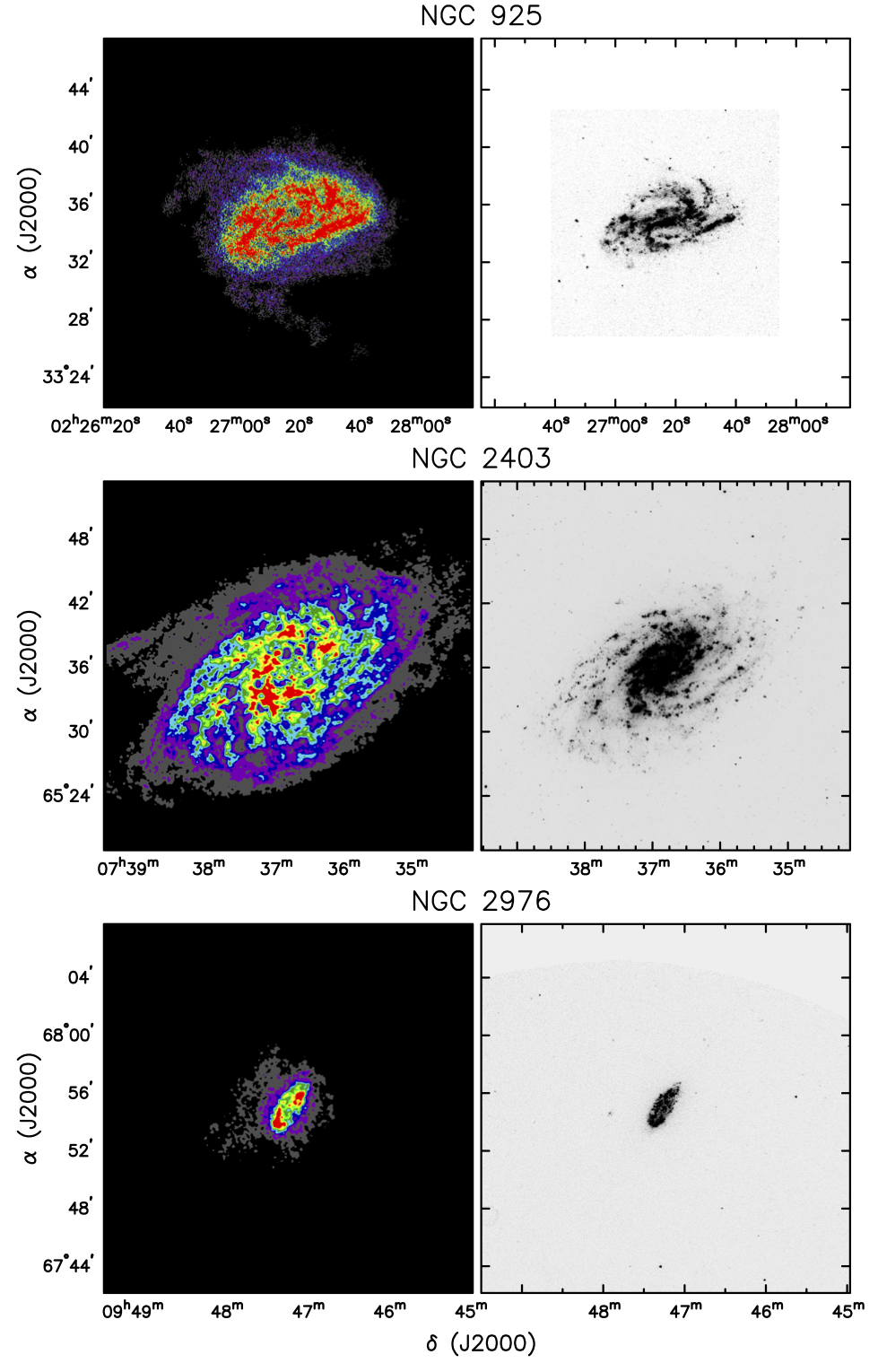}
\caption{Galaxy images; north is up and east is left. Left: \ion{H}{1} column density image from THINGS. Right: $GALEX$ FUV image.\label{galimages}}
\end{figure}
\begin{figure}[t!]
\epsscale{0.8}
\centering
\plotone{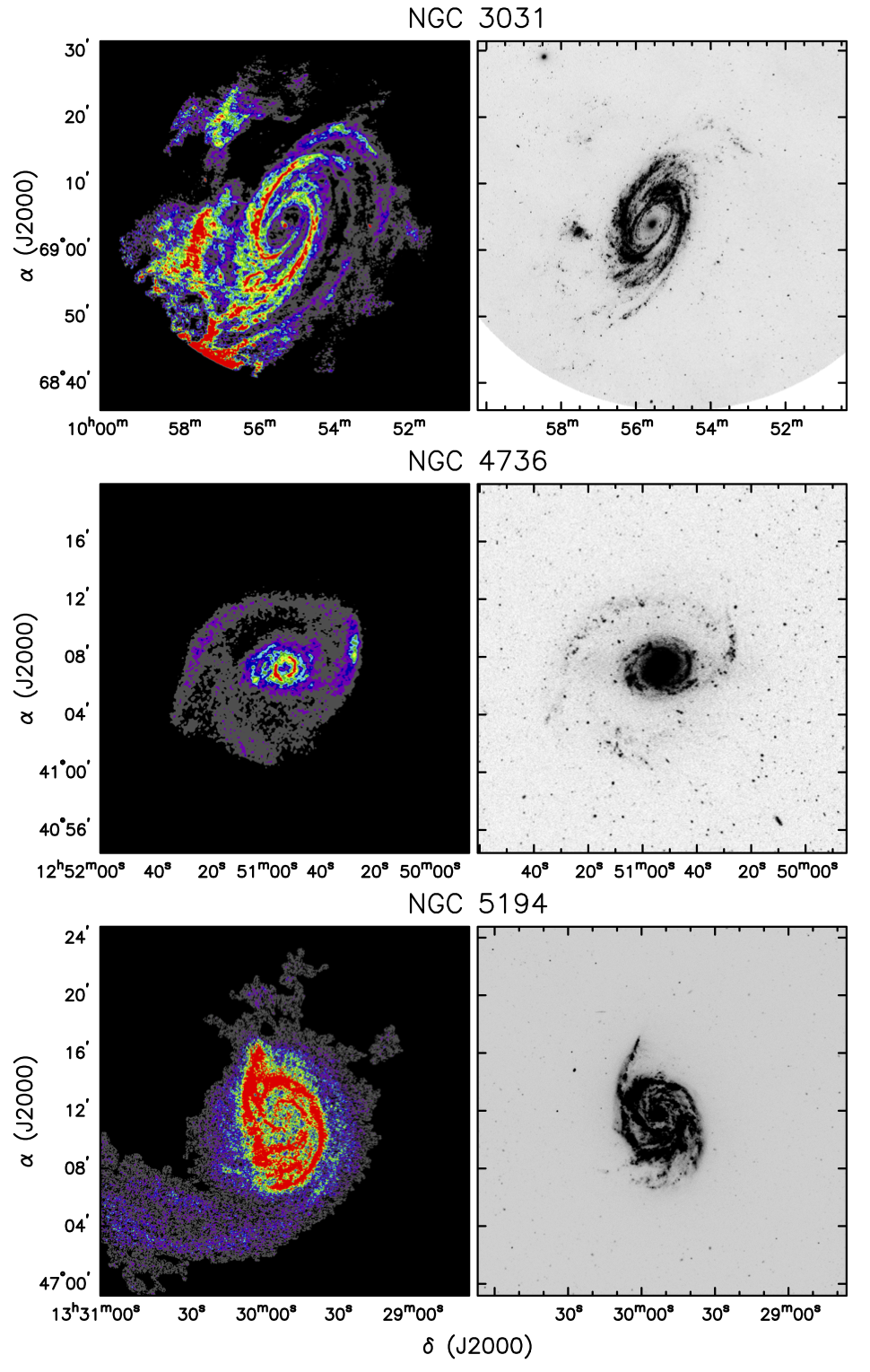}\\
Figure~\ref{galimages} continued.
\end{figure}
\begin{figure}[t!]
\epsscale{0.8}
\centering
\plotone{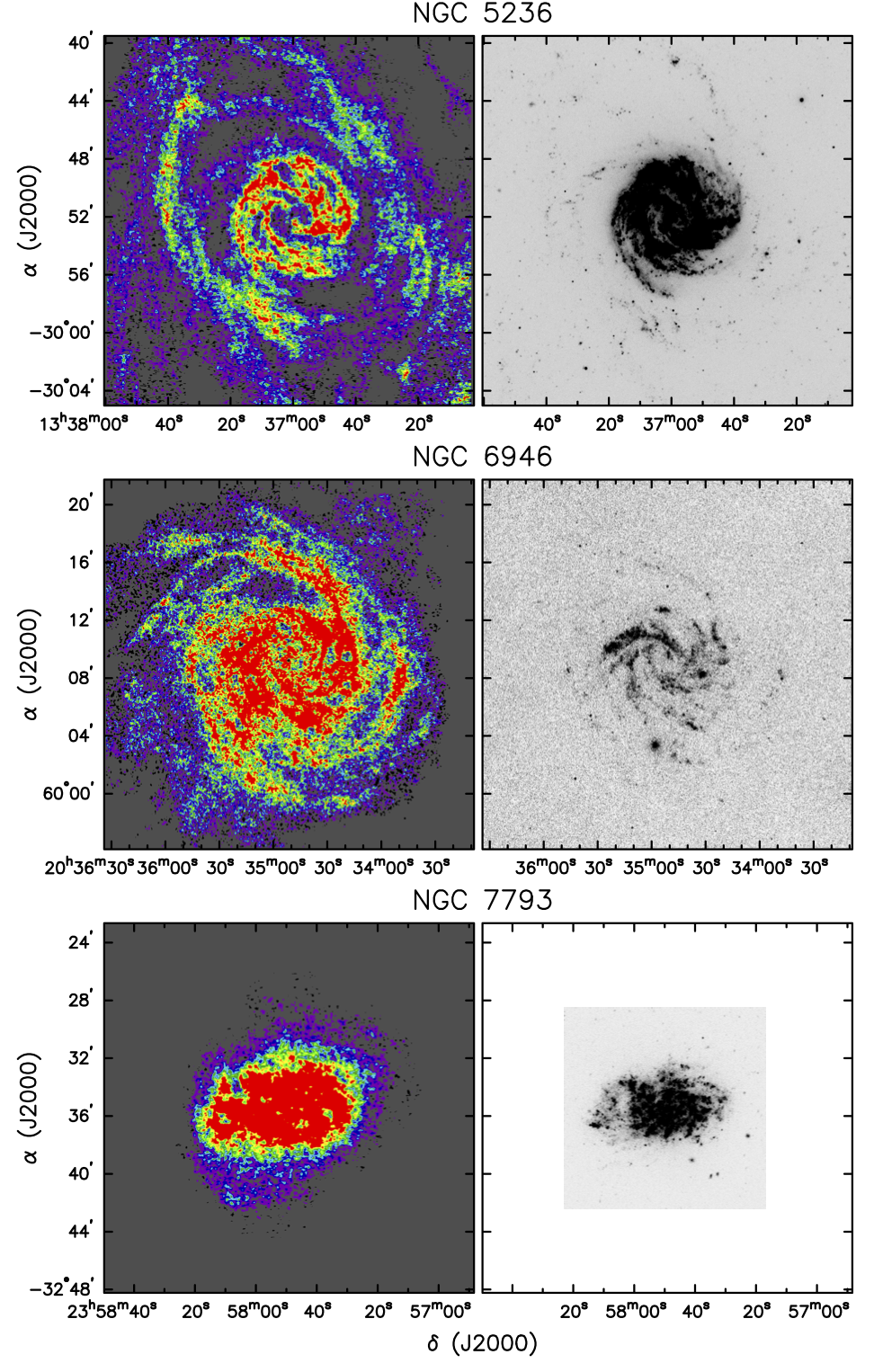}\\
Figure~\ref{galimages} continued.
\end{figure}
\begin{figure}[t!]
\epsscale{0.8}
\centering
\plotone{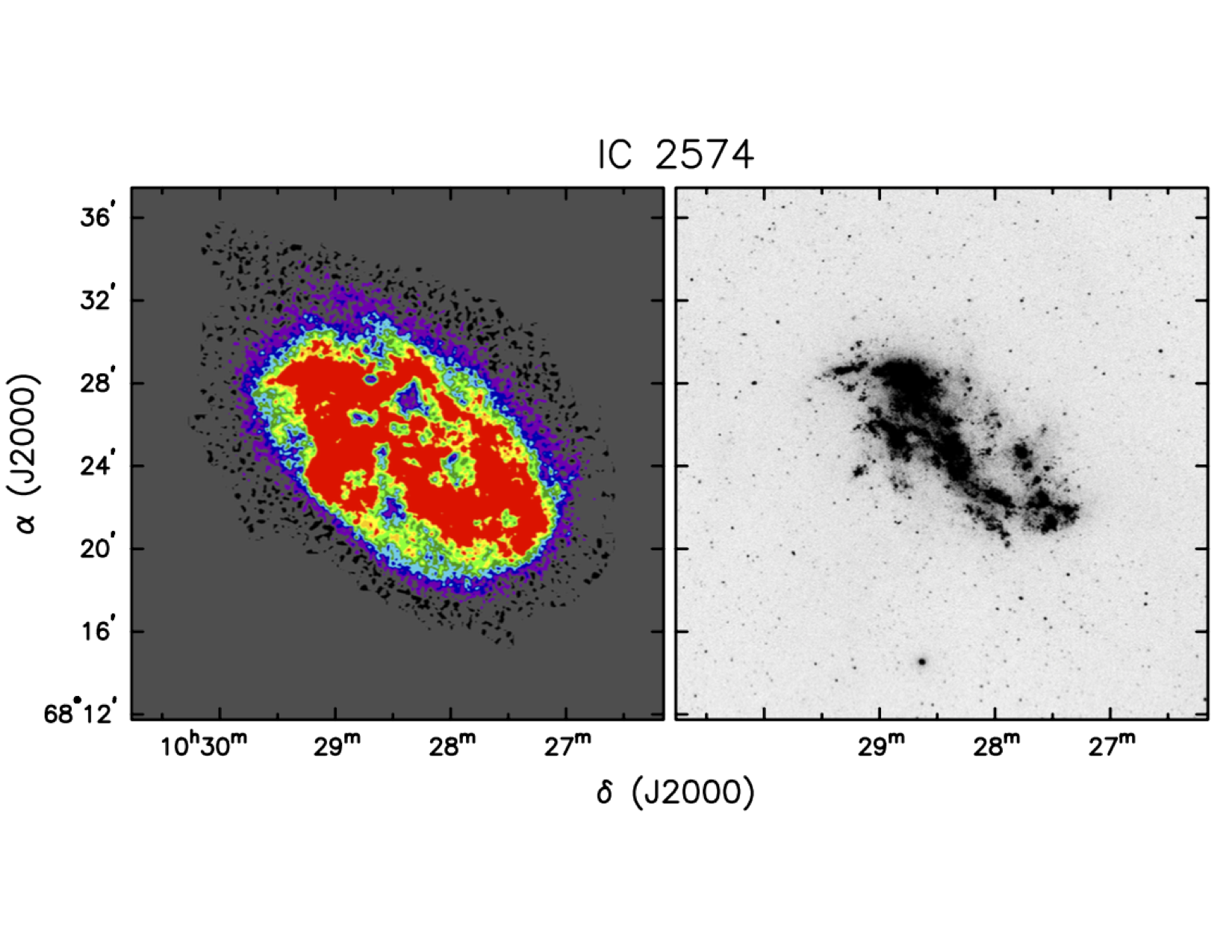}\\
Figure~\ref{galimages} continued.
\end{figure}

Figure~\ref{galimages} shows both the \ion{H}{1} column density map and FUV images of all the galaxies. Table~\ref{galobsdata} summarizes the image parameters for each galaxy. In the final analysis, the \ion{H}{1} and FUV images are aligned and trimmed to be the same size, listed in the last column in Table~\ref{galobsdata}. 


\begin{deluxetable}{lccr}
\tablecaption{Parameters of \ion{H}{1} and FUV Images\label{galobsdata}}
\tablewidth{0pt}
\tabletypesize{\normalsize}
\tablehead{\colhead{} & \colhead{\ion{H}{1} Scale} & \colhead{FUV Scale}  & \colhead{Final Size} \\
\colhead{Galaxy} & \colhead{(\arcsec\,pix$^{-1}$)} & \colhead{(\arcsec\,pix$^{-1}$)} & \colhead{(pix)}  
}

\startdata

NGC 925 & 1.5 & 1.5  & 629 $\times$ 629\\
NGC 2403  & 1.0 & 1.5 & 2000 $\times$ 2000 \\
NGC 2976  & 1.5 &1.5 &1024  $\times$ 1024 \\
NGC 3031 & 1.5 & 1.5 & 2000 $\times$ 2000\\
NGC 4736 & 1.5 & 1.5 & 1024  $\times$ 1024\\
NGC 5194 & 1.5 & 1.5  & 1024 $\times$ 1024\\
NGC 5236 & 1.5\tablenotemark{a} &1.5 &1024 $\times$ 1024\\
NGC 6946 & 1.5 & 1.5 & 1024 $\times$ 1024\\
NGC 7793 & 1.5 & 1.5 & 557 $\times$ 557\\
IC 2574 & 1.5 & 1.5 & 1024 $\times$ 1024\\
\enddata
\tablenotetext{a}{Note that the image scale is found to be different from \citet{Walter}.}

\end{deluxetable}

\section{ANALYSIS \label{analysis}}
\subsection{Higher Order Statistical Moments \label{HOS}}
The first through fourth statistical moments of a given distribution are the mean, variance, skewness, and kurtosis. In order to compare with the results from \citet{Burk}, we adopted their {\it standard score} scaling method for the \ion{H}{1} column density distribution:
\begin{equation}
\xi = \frac{\xi_{o} - \bar\xi_{o}}{\sigma_{o}},
\end{equation}
where $\xi$ is the new normalized value, $\xi_{o}$ is the original column density value, $\bar\xi_{o}$ is the arithmetic mean, and $\sigma_{o}$ is the standard deviation of the original data set. Note that we only consider pixels with column density value $\xi_{o}$ greater than zero. This new normalized value represents the difference between the original value and the mean, in units of the standard deviation.  

Following this scaling operation, for an image with the column density value per pixel \emph{i} given by $\xi_i$, the mean $\bar\xi$ and variance $\sigma^2_\xi$ can be calculated as
\begin{equation}
\bar\xi= \frac{1}{N} \sum \limits_{i=1}^N \xi _i 
\end{equation}
and
\begin{equation}
\sigma^2_\xi = \frac{1}{N - 1} \sum\limits_{i=1}^N \left(\xi _i - \bar{\xi} \right)^2 
\end{equation}
respectively, where $N$ is the total number of pixels (with $\xi_{o}$ values $>0$) in the image. 

The HOS moments--- skewness ($\gamma_\xi$) and kurtosis ($\beta_\xi$)--- can then be calculated. The third statistical moment, skewness, is defined as
\begin{equation}
\gamma _\xi = \frac{1}{N} \sum\limits_{i=1}^N \left(\frac{\xi _i - \bar{\xi}}{\sigma _\xi}\right)^3 .
\end{equation}
When skewness does not equal to zero, the distribution of the data are skewed and have an elongated tail ($i.e.$ a Gaussian distribution has a skewness of zero, and the mean and the mode coincide at the same location). If the skewness value is positive, the distribution has its peak (or the mode) leftwards of a normal Gaussian and a tail extended rightwards, and vice versa for a negative skewness value. A distribution with a more positive skewness value has its peak shifted further leftwards. When applied to our \ion{H}{1} column density images, a distribution with positive skewness represents that the sample data is skewed towards lower density values, with a small number of pixels having high density values. The standard error in skewness (SES) can be estimated by $\sqrt{6/N}$ \citep{Burk,tf96}.

The fourth statistical moment, kurtosis, is defined as
\begin{equation}
\beta _\xi = \frac{1}{N}\sum\limits_{i=1}^N \left(\frac{\xi _i - \bar{\xi}}{\sigma _\xi}\right)^4 - 3.
\end{equation}

Kurtosis is a measure of the ``sharpness" or ``flatness'' of a distribution's peak as compared to a Gaussian distribution ($i.e.$ again a Gaussian distribution has a kurtosis of zero). A positive kurtosis indicates a distribution with a sharper peak and elongated tails, while a negative kurtosis indicates a distribution with a flatter peak and non-elongated tails. The standard error in kurtosis (SEK) can be estimated by $\sqrt{24/N}$ \citep{Burk,tf96}. 

As mentioned in Section~\ref{intro}, both skewness and kurtosis have a dependance on turbulent properties, specifically the sonic Mach number, $\mathcal{M}_{s}$ \citep{Burk}. To better understand the interpretations from the results of \citet[][specifically their Figures~3 and 5]{Burk}, we summarize the relations below. When kurtosis has a value of $\beta \sim 0$ (and skewness of $\gamma \sim 0.7$), $\mathcal{M}_{s} = 1$. When $\beta < 0$ (and $\gamma < 0.7$), $\mathcal{M}_{s} < 1$ and both moments have a flat dependence on $\mathcal{M}_{s}$. Thus when gas motion is in subsonic regime, turbulent compression is not strong and has a weak correlation with the gas velocity. 

However, when $0 < \beta \lesssim 3$, gas motion can be subsonic or supersonic, or so-called transonic. One way to determine the gas motion in the regions of study in a galaxy, is to examine the correlation between the two moment values in these regions (see Section~\ref{kernel} and Figure~\ref{svsk} below). If overall $\beta$ values have a single dependence on $\gamma$ values, gas motion is supersonic; if $\beta$ values do not have a single correlation with $\gamma$ values in these regions, gas motion is subsonic. In other words, some regions in a galaxy can have the same $\beta$ value but different $\gamma$ values (here $-1 < \gamma <1.5$) when gas motion is subsonic. When $\beta > 3$, the ambiguity vanishes and gas motion is all supersonic. Kurtosis and skewness both then have a strong correlation with $\mathcal{M}_s$ when $\beta > 3$, and so the higher gas velocity in their supersonic motion, the stronger the compression. However, care has to be taken when interpreting the sources of such compression. \citet{Burk} suggested that the gas compression can be due to turbulence but not necessarily, it can also be due to gas shearing or just local shocks. Such compression may not also be directly related to star forming regions.

\subsection{Whole Image HOS Moments \label{whole}}
\begin{figure}[t!]
\epsscale{1.0}
\centering
\plotone{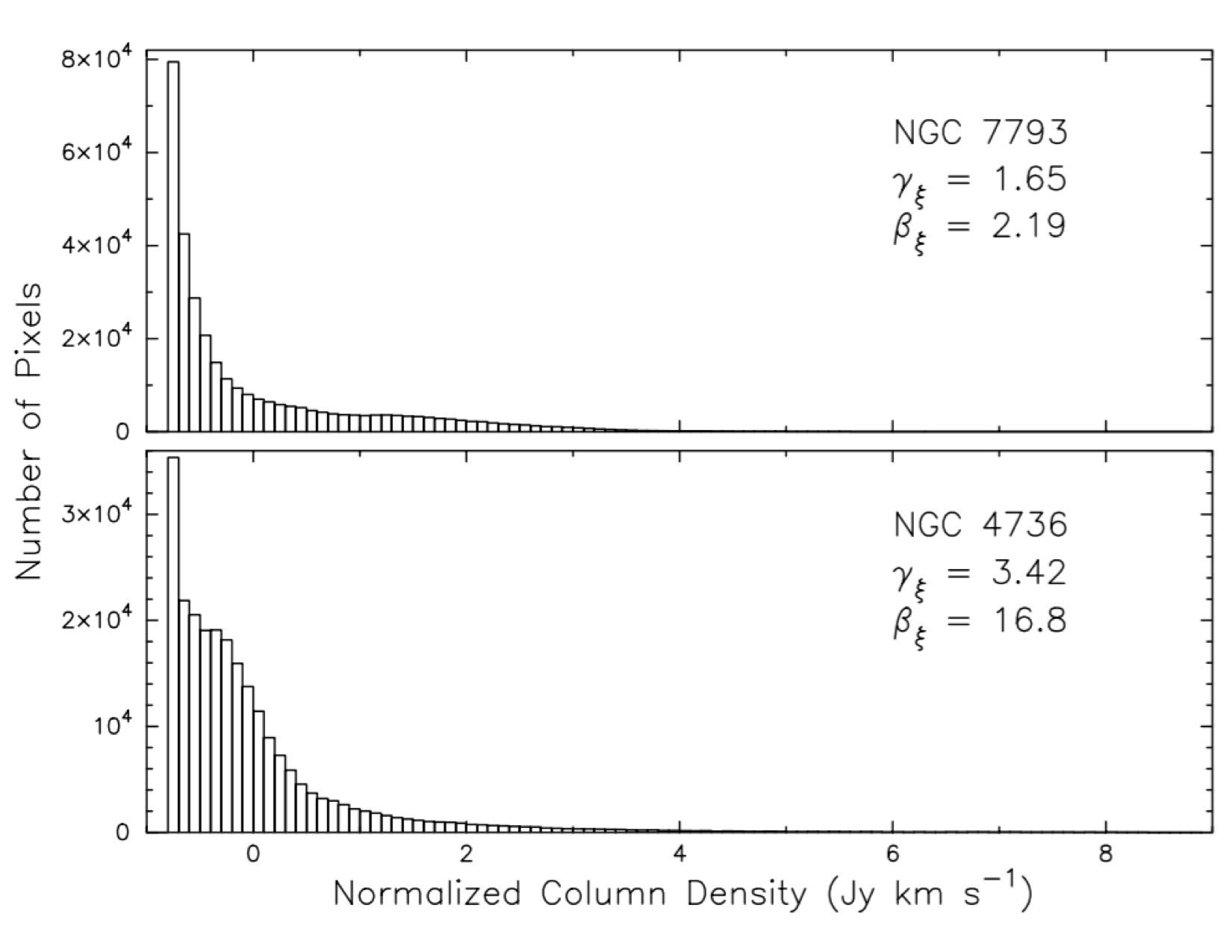}
\caption{The distribution of normalized column density in NGC~7793 and NGC~4736, from $-1.0$ to $9.0$ in $100$ bins. Note that the vertical scale is different. Their HOS values, skewness $\gamma_\xi$ and kurtosis $\beta_\xi$, are also shown.\label{galhist}}
\end{figure}

Skewness and kurtosis were first calculated for each \ion{H}{1} image as a whole, and the results for each galaxy are summarized in Table \ref{galstat}. All galaxies in our sample have positive skewness and kurtosis values, and the distributions of their normalized column densities are highly non-Gaussian and skewed towards the low column density values with a much elongated tail. Based on \citet{Burk}, these values suggest that overall gas motions in all galaxies are in the supersonic regime.  NGC~7793 has the lowest kurtosis ($\beta_\xi =2.19$) while NGC~4736 has the highest kurtosis ($\beta_\xi=16.8$) among the galaxies. Their distributions and HOS values are shown in Figure~\ref{galhist}. Notice how the shapes of the distribution are different which demonstrates the different kurtosis values. As expected, \ion{H}{1} gas is distributed unevenly in galaxies with concentrations and low density regions. Comparing to their images, NGC~4736 shows a high concentration of gas in the central regions with relatively low \ion{H}{1} density in the arms. On the other hand, NGC~7793 shows relatively evenly distributed \ion{H}{1} gas.


\begin{deluxetable}{lccrcr}
\tablecaption{Statistical Moments of \ion{H}{1} images\label{galstat}}
\tablewidth{0pt}
\tabletypesize{\normalsize}
\tablehead{\colhead{} & \colhead{Mean\tablenotemark{a}} & \colhead{St. Dev.\tablenotemark{b}} & \colhead{Normalized Mean} & \colhead{Skewness} & \colhead{Kurtosis}\\
\colhead{Galaxy} & \colhead{($\bar\xi_{o}$)} & \colhead{($\sigma_{o}$)} & \colhead{($\bar\xi$)} & \colhead{($\gamma_{\xi}$)} & \colhead{($\beta_{\xi}$)}
}

\startdata

NGC 925 & 689 & 804 & 5.45 $\times\  10^{-16}$ & 1.56 & 2.24 \\
NGC 2403 & 513 & 553 & $-$3.01 $\times\  10^{-14}$ & 1.65 & 3.23 \\
NGC 2976 & 235 & 454 & 2.16 $\times\  10^{-16}$ & 3.54 & 14.57 \\ 
NGC 3031 & 585 & 673& $-$3.53 $\times\  10^{-15}$ & 1.94 & 4.86 \\
NGC 4736 & 186 & 239 & 9.21 $\times\  10^{-16}$ & 3.42 & 16.80 \\
NGC 5194 & 311 & 440  & 3.52 $\times\  10^{-15}$ & 2.31 & 6.08 \\
NGC 5236 & 224 & 220 & $-$6.03 $\times\  10^{-15}$ & 1.92 & 4.99 \\
NGC 6946 & 399 & 400 & $-$7.33 $\times\  10^{-15}$ & 1.63 & 3.38 \\
NGC 7793 & 444 & 568& 1.07 $\times\  10^{-14}$ & 1.65 & 2.19 \\
IC 2574  & 577 & 625 & 3.54 $\times\  10^{-15}$ & 1.57 & 3.05 \\

\enddata

\tablenotetext{a}{Mean of the original column density. Note only values $>0$ are included.}
\tablenotetext{b}{Standard deviation of the data included in calculating $\bar\xi_{o}$.}
\end{deluxetable}

\begin{figure}[t!]
\epsscale{1.0}
\centering
\plotone{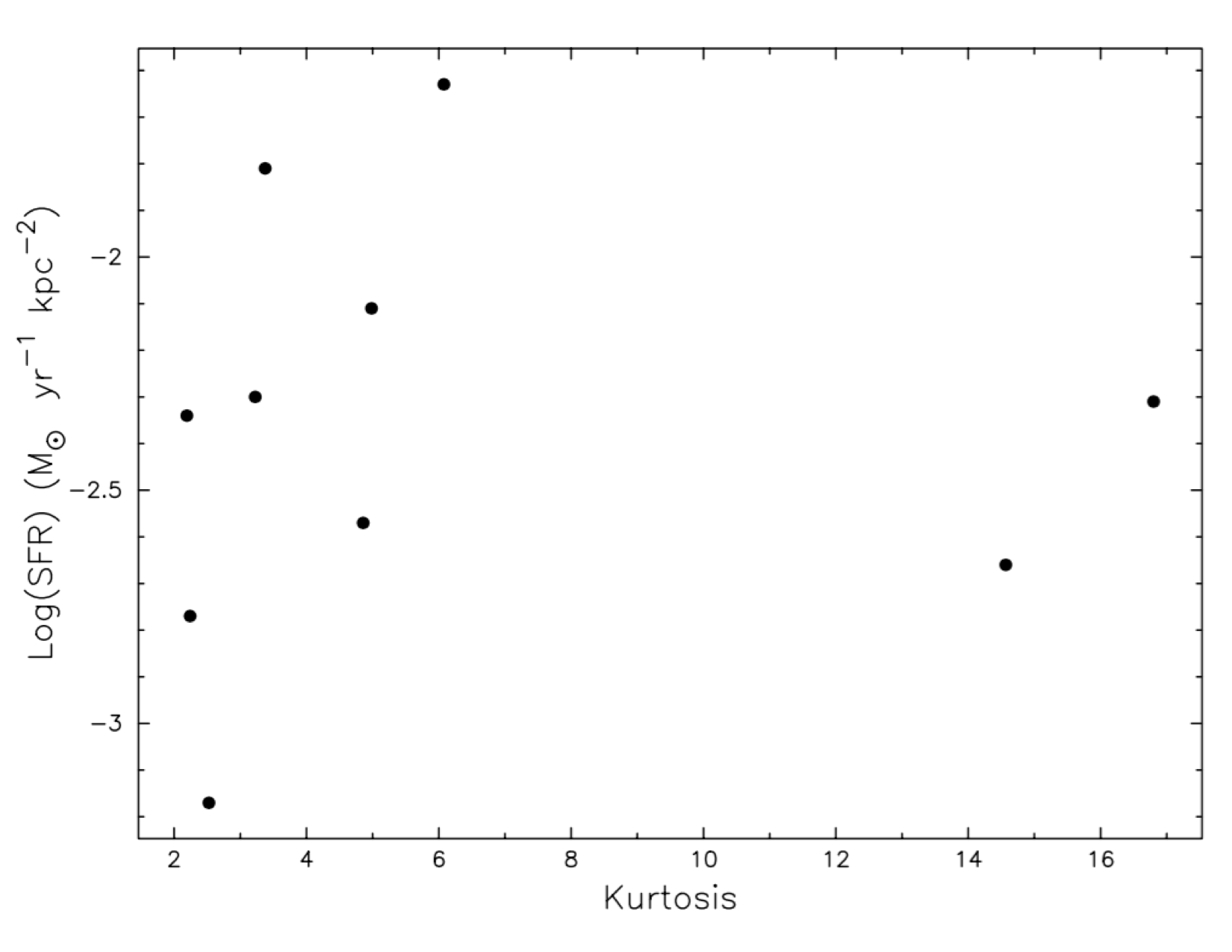}
\caption{Log(SFR) vs. Kurtosis for each galaxy as a whole. Star formation rate, normalized by the area of the galaxy, is in units of $\solarmass$ yr$^{-1}$ kpc$^{-2}$.\label{kvssfr}}
\end{figure}

We also compare their kurtosis values with the log of star formation rate (SFR) listed in Column~(10) in Table~\ref{galprop}, normalized by their area. Here we estimate the area of a galaxy using its optical distance $D$ and radius $R_{25}$, listed in Columns~(4) and (5) in Table~\ref{galprop}. The result is shown in Figure~\ref{kvssfr}, 
however, there is no evident correlation between kurtosis in a galaxy as a whole and its global SFR--- for similar kurtosis values in galaxies, there can be more than an order of magnitude difference in the normalized SFR. 

\subsection{HOS Moments as a Function of Radius \label{annuli}}

Our next application of the HOS was their calculation as a function of radius from the center of each galaxy. We chose $20$ pixels as the central radius for galaxy images with a pixel scale of $1.5\arcsec$. This size was chosen so that at least $50$ independent beam sizes were included in each annulus to ensure a significant signal-to-noise ratio. The annuli continued outwards in $20$-pixel increments to just beyond the edge of each galaxy. For NGC~2403, with a pixel scale of $1.0\arcsec$, we use $30$ pixels for the central radius as well as the increment of the annuli. Each of these annuli extends approximately $30\arcsec$~in width, or approximately $730$~pc at a distance of $5.0$~Mpc.

In order to calculate the statistics in each annulus, first we calculated the apparent distance $s$ of each pixel in the image based on its coordinate ($x, y$) relative to the galactic center ($x_{o}, y_{o}$) listed in Column~(2) and (3) in Table~\ref{galprop}. We then corrected for the galaxy's geometry and derived the radius in the plane of the galaxy of each pixel using the formula
\begin{equation}
r = s [\sec^2 i - \tan^2 i~\cos^2 (\eta - \phi)]^{1/2},
\end{equation}
where $i$ and $\phi$ are the inclination and position angle of the galaxy, listed in Columns~(6) and (7) in Table~\ref{galprop}, and $\eta$ is the angle between north and the pixel measured from ($x_{o}, y_{o}$). The angle $\eta$ does not need to be determined, since $\cos^2(\eta - \phi)$ can be calculated knowing that $\tan(\eta - \phi) = (y - y_{o}) / (x - x_{o})$. Once $r$ was determined, each pixel was assigned to an annulus, and the HOS moments for each annulus were calculated. As an example, these statistics for NGC~5194 are shown in Table~\ref{n5194stat}.


\begin{deluxetable}{lrcrcc}
\tablecaption{HOS as a Function of Radius for NGC~5194\label{n5194stat}}
\tablewidth{0pt}
\tabletypesize{\normalsize}
\tablehead{\colhead{Annulus} & \colhead{Pixels} & \colhead{Skewness} & \colhead{Kurtosis} & \colhead{} & \colhead{}\\
\colhead{(pix)} & \colhead{included\tablenotemark{a}} & \colhead{($\gamma_{\xi}$)} & \colhead{($\beta_{\xi}$)} & \colhead{SES\tablenotemark{b}} & \colhead{SEK\tablenotemark{b}}
}

\startdata

0 --- 20 & 936 & 0.31 & $-$0.59 & 0.080 & 0.160 \\
20 --- 40 & 2800 & 0.65 & $-$0.03 & 0.046 & 0.093 \\
40 --- 60  & 4655 & 0.88 & 1.23 & 0.036 & 0.072 \\
60 --- 80  & 6534 & 1.05 & 1.57 & 0.030 & 0.061 \\
80 --- 100  & 8383 & 0.80 & 0.62 & 0.027 & 0.054  \\
100 --- 120  & 10199 & 0.87 & 0.52 & 0.024 & 0.049 \\
120 --- 140  & 11841 & 1.41 & 2.10 & 0.023 & 0.045  \\
140 --- 160 & 13434 & 1.44 & 2.22 & 0.021 & 0.042  \\
160 --- 180 & 15033 & 1.84 & 3.85 & 0.020 & 0.040 \\
180 --- 200 & 15838 & 2.34 & 7.31 & 0.019 & 0.039  \\
200 --- 220 & 16024 & 2.11 & 5.27 & 0.019 & 0.039  \\
220 --- 240 & 14932 & 2.04 & 4.40 & 0.020 & 0.040 \\
240 --- 260 & 14493 & 2.17 & 5.04 & 0.020 & 0.041  \\
260 --- 280 & 13000 & 2.17 & 5.12 & 0.021 & 0.043 \\
280 --- 300 & 11714 & 2.46 & 6.60 & 0.023 & 0.045  \\
300 --- 320 & 9835 & 2.98 & 12.15 & 0.025 & 0.049  \\
320 --- 340 & 9276 & 1.60 & 2.39 & 0.025 & 0.051 \\
340 --- 360 & 8251 & 1.84 & 4.10 & 0.027 & 0.054  \\

\enddata
\tablenotetext{a}{Total number of pixels (with values $>$ 0 before normalization) included within the annulus.}
\tablenotetext{b}{Standard error in skewness and kurtosis, calculated based on \citet{Burk}.}
\end{deluxetable}

\begin{figure}[t!]
\epsscale{1.0}
\plotone{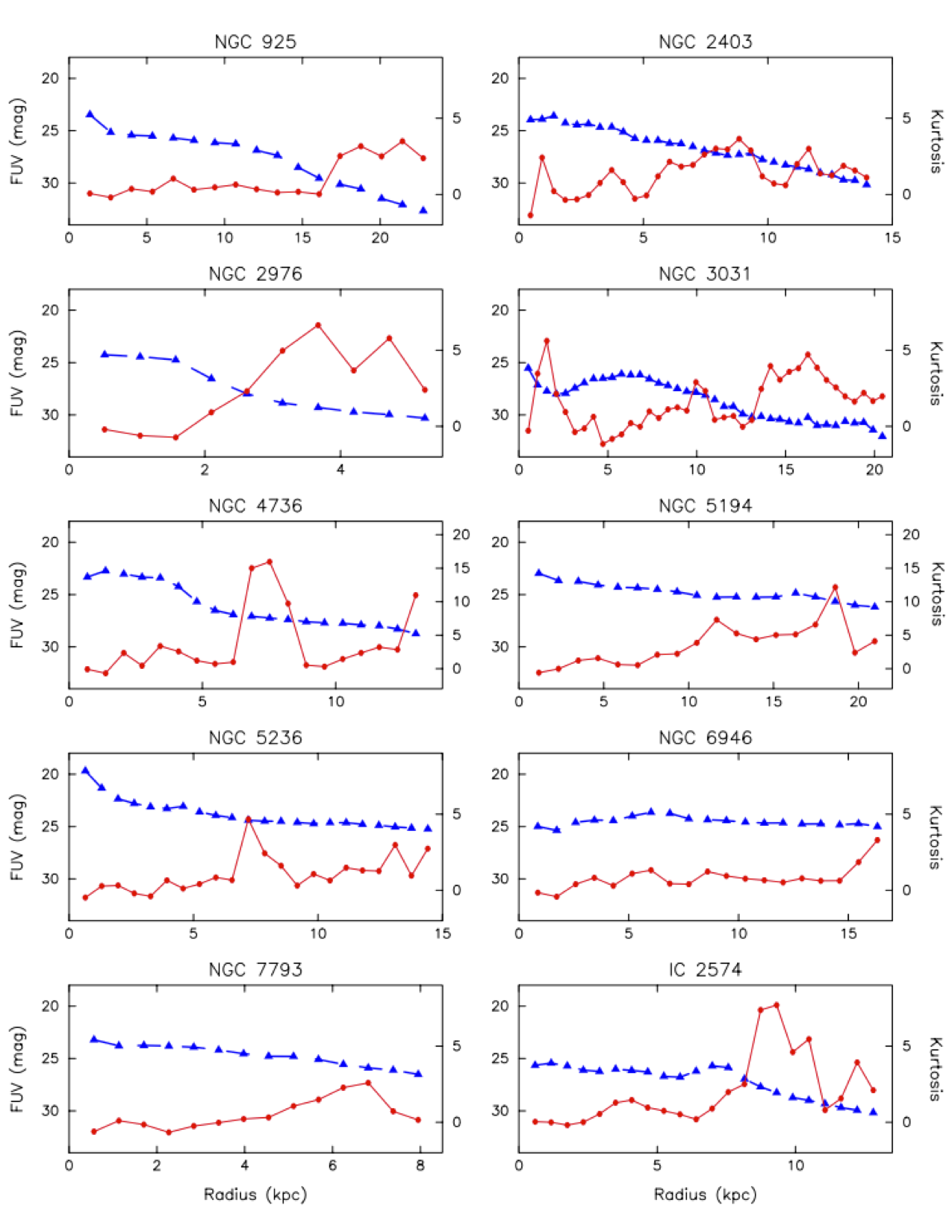}
\centering
\caption{Kurtosis, plotted on the right axis, and FUV surface brightness (mag), plotted on the left axis, as a function of radius (kpc) for each galaxy. Blue triangles show the FUV magnitudes at each annulus connected by the dashed line. Red circles show the kurtosis values connected by the solid line. Typical errors in FUV and kurtosis are in the order of $10^{-1}$ mag and $10^{-2}$ respectively.\label{kvsfuv}}
\end{figure}

In order to compare with the star formation rate in each annulus, the IRAF\footnote{IRAF is distributed by the National Optical Astronomy Observatory, which is operated by the Association of Universities for Research in Astronomy (AURA) under a cooperative agreement with the National Science Foundation.} {\sc ellipse} function was used to perform surface photometry in concentric ellipses on the FUV images, matching those created for the \ion{H}{1} images. We then parsed the data from those ellipses into separate annuli and constructed FUV magnitudes per annulus.
Figure~\ref{kvsfuv} shows the comparison of kurtosis and FUV surface brightness as a function of radius (in~kpc) for each galaxy. 

As expected, most of the FUV surface brightnesses show a general downward trend for larger annuli. NGC~6946, however, has a relatively flat brightness across the galaxy. In some galaxies, such as NGC~3031 and IC~2574, there is an increase in brightness at mid-radius regions, most likely due to crossing from a dimmer part of a galaxy to a brighter one. 

On the other hand, the kurtosis, with values between $-1$ and $8$, appears to have a loosely upward trend in most galaxies, however the progression from central annuli to outermost was noisy and in general showed no correlation with fluctuations in the FUV surface photometry. The highest kurtosis value occurs in NGC~4736 ($\beta_{\xi} \sim 16$) at radii between $6$ to $8$~kpc. This is most likely due to the concentrated \ion{H}{1} region in the western arm (see Figure~\ref{galimages}), which contributes to a much elongated tail towards the high density values in the distribution, and therefore gives a high kurtosis value at such radii. In general, we do not find any correlation between FUV surface brightness and kurtosis among the galaxies in this analysis.

\subsection{HOS Moments in a Grid \label{kernel}}

Our final application of the HOS was to calculate them in a grid of $32$~square pixels ($48$ square pixels for NGC~2403) across the image of each galaxy. At a distance of $5.0$~Mpc, $32$~pixels is equal to $\sim 48\arcsec$, or $\sim1$~kpc. This method is in analogy to \citet{Burk}, where they generated HOS maps using a circular moving kernel of $35$~pixels in radius. First, we aligned and trimmed the \ion{H}{1} and FUV images to the same size depending on which was smaller. The final sizes are listed in the last column in Table~\ref{galobsdata}. We then split each image into a grid, calculated the HOS in each square, and created the HOS maps for each galaxy. We masked the pixels in the background of each image to a value of $0$, and squares along the edges of a galaxy could include a number of these zero-valued pixels. In order to correctly calculate the HOS in these squares, we developed a cutoff where a square should contain at least a certain number of pixels, $N_{cutoff}$, with values $>0$. In other words, if the number of pixels with values $>0$ is less than $N_{cutoff}$ in a square, the HOS were not calculated and were assigned zero. For skewness calculation, $N_{cutoff} = 192$ pixels and for kurtosis, $N_{cutoff} =  384$ pixels.  These cutoffs correspond to a SES $\sim0.2$ and SEK $\sim0.25$ respectively. 

\begin{figure}[t!]
\epsscale{0.9}
\centering
\plotone{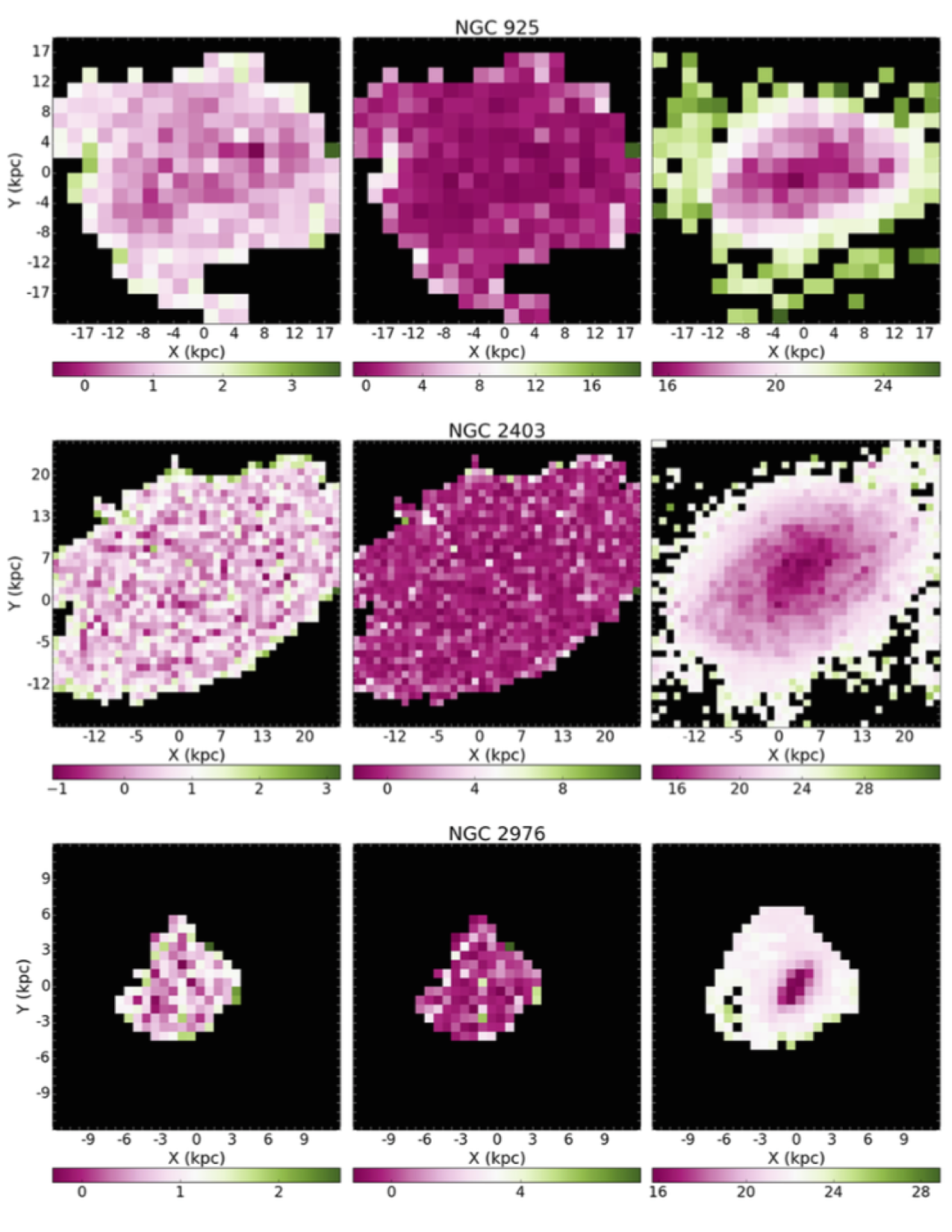}
\caption{Left to right: skewness, kurtosis, and FUV (mag) maps of each galaxy with color bars shown. Note that each square in NGC~2403 FUV map is 32~pixels wide, which gives the same physical size of 48$\arcsec$ of each square in the skewness and kurtosis maps.\label{gridmaps}}
\end{figure}

In order to make a grid-to-grid comparison with the FUV images, we then divided the FUV images into the same grid and calculated the total FUV magnitude for each square. Note that the original FUV images of NGC~2976 and NGC~3031 were limited by the instrument, causing anomalous edges of squares in the grid images. In order to show the direct comparison to the skewness and kurtosis, a mask similar to the distribution in the skewness map is applied to the grid image for each galaxy. Skewness, kurtosis, and FUV grid maps for all the galaxies are shown in Figure~\ref{gridmaps}.

\begin{figure}[t!]
\epsscale{0.9}
\centering
\plotone{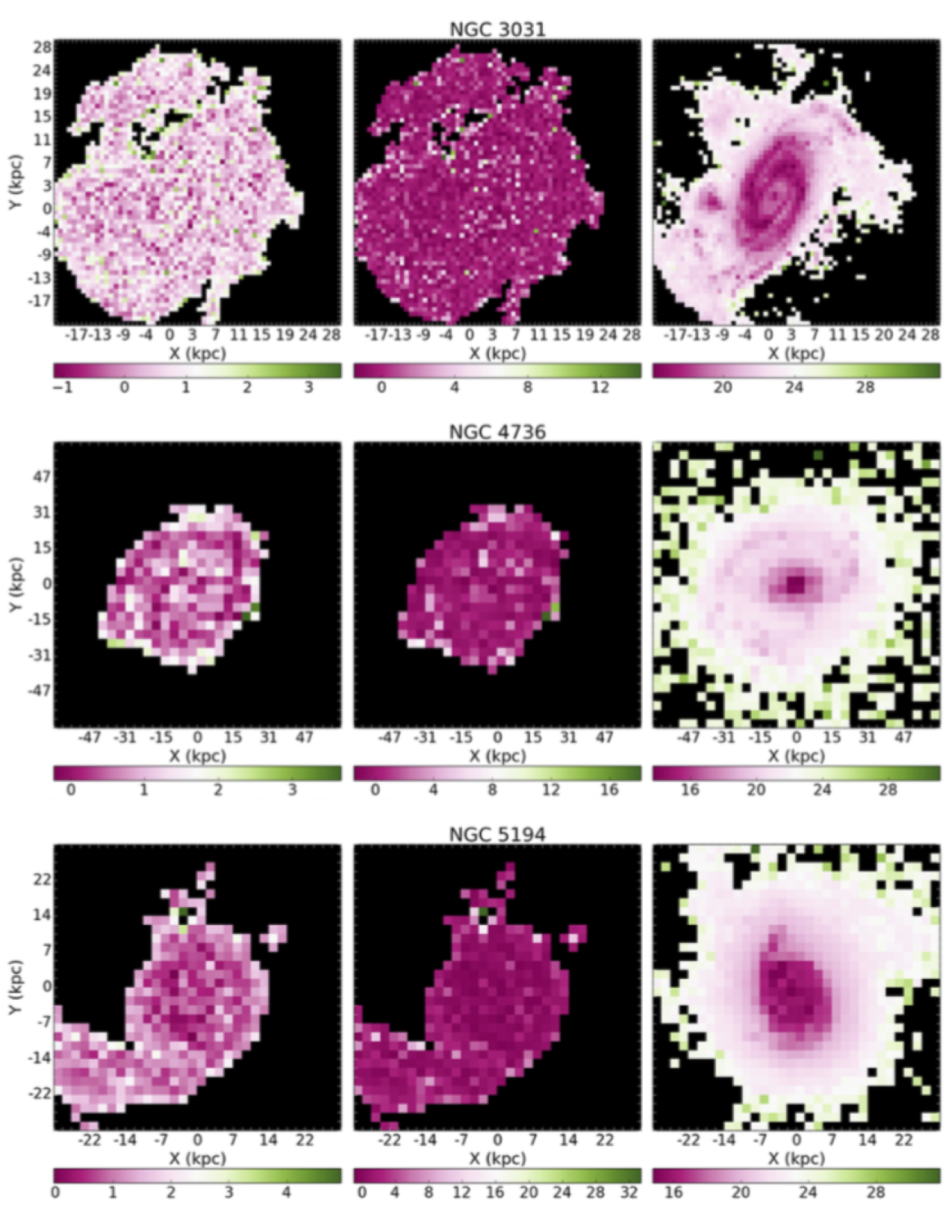}\\
Figure~\ref{gridmaps} continued.
\end{figure}
\begin{figure}[t!]
\epsscale{0.9}
\centering
\plotone{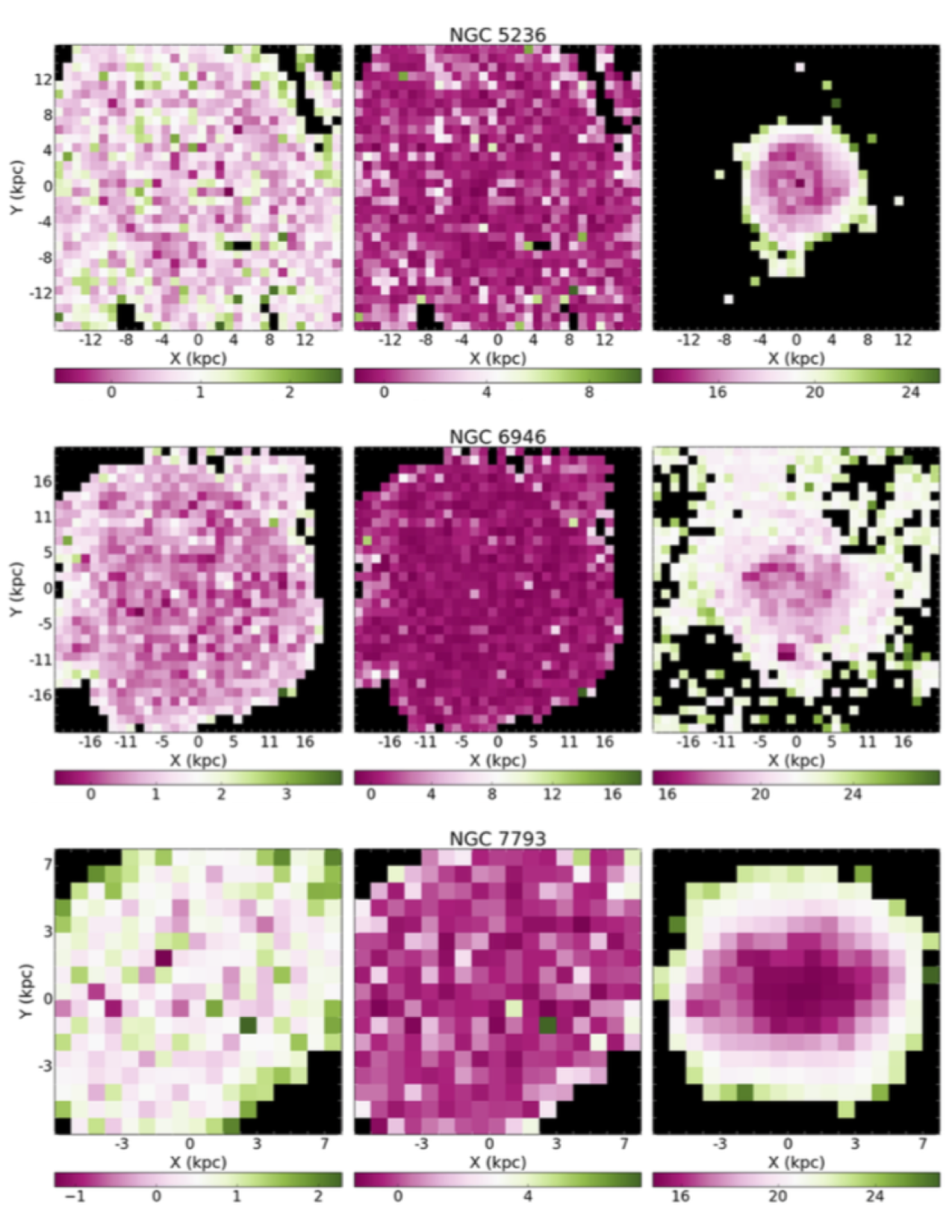}\\
Figure~\ref{gridmaps} continued.
\end{figure}
\begin{figure}[t!]
\epsscale{0.9}
\centering
\plotone{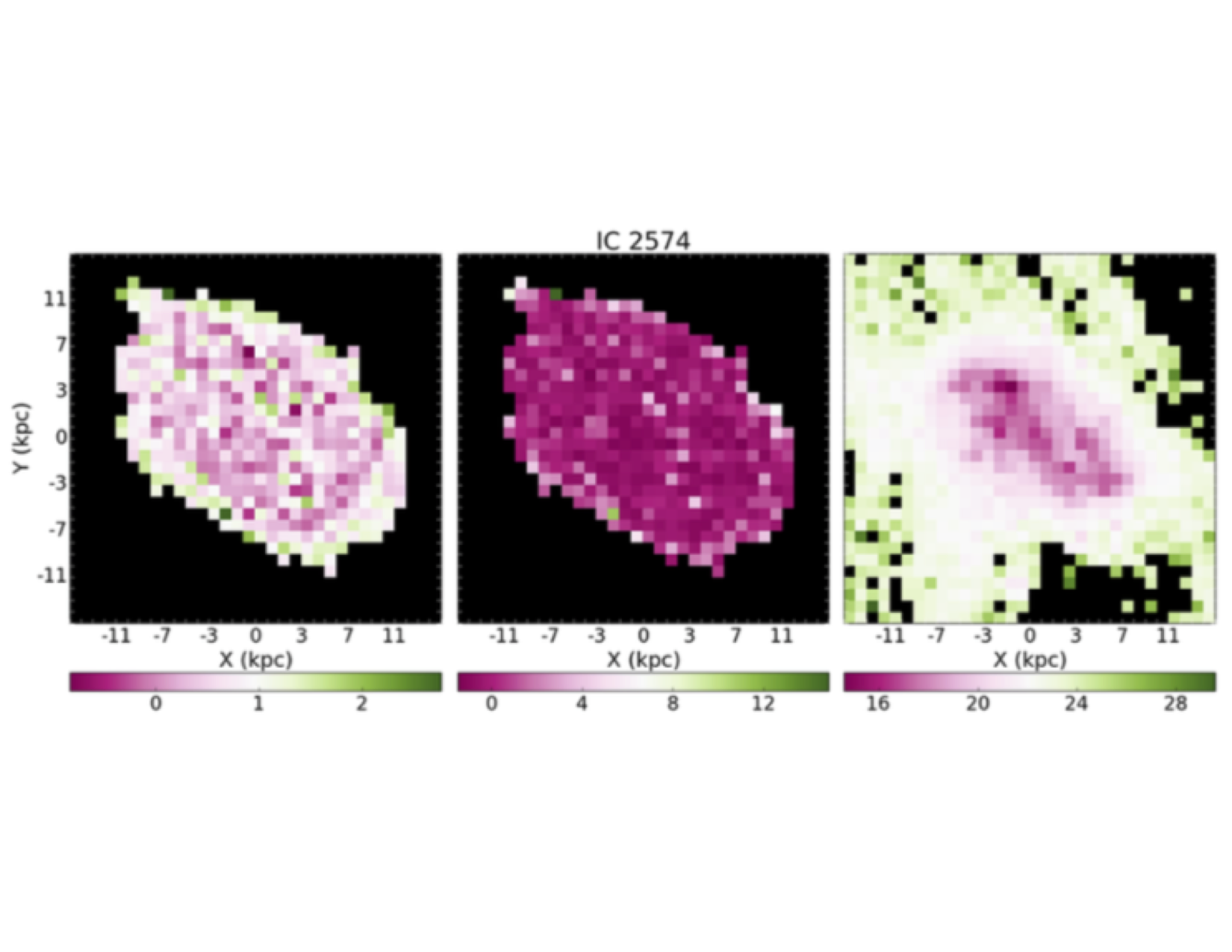}\\
Figure~\ref{gridmaps} continued.
\end{figure}

These maps show that skewness generally appears to increase with distance from the center of the galaxy, and in some galaxies, for example NGC~2403, NGC~4736, NGC773, high skewness values occur around the edges. On the other hand, kurtosis remains relatively uniform across each galaxy with some fluctuations around the edges. Comparing to the FUV grid maps, skewness maps appear to trace some galaxy arms, especially in NGC~3031 and NGC~4736, but no patterns relative to star-forming regions are found in the kurtosis maps. 

\begin{figure}[t!]
\epsscale{1}
\centering
\plotone{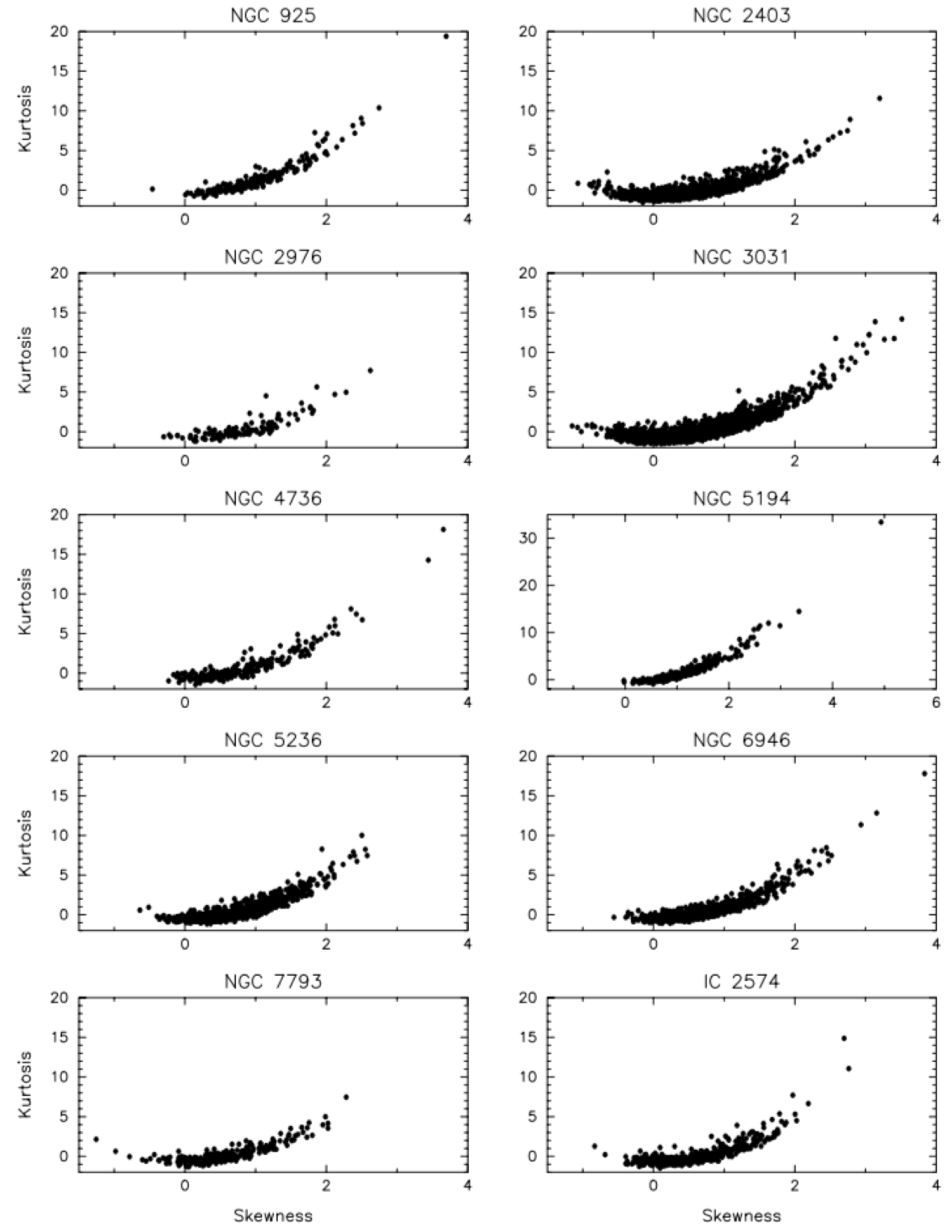}
\caption{Kurtosis versus skewness in each kernel of the grid for each galaxy. Note that the axes for NGC~5194 are different from the rest of the plots.\label{svsk}}
\end{figure}

Figure~\ref{svsk} shows the grid-to-grid relation of kurtosis and skewness for each galaxy. Similar to what we find in the analysis of whole image HOS moments (Section~\ref{whole}), the moments show a good correlation and generally match to the supersonic result of \citet{Burk}, indicating that the gas motions in these grids are supersonic. However as noted earlier in Section~\ref{HOS}, such supersonic compressions may not directly due to turbulent motions. In our samples, though we do not rule out the possibility of these compressions due to turbulence, it is likely that these compressions are due to local shearing or shocks.

\section{DISCUSSION AND SUMMARY \label{discuss}}

\subsection{Correlation Between Turbulence and Star Formation}

Studies have suggested that turbulence in ionized gas regions is to be expected, and that one can see a correlation between turbulence and star formation \citep[e.g.][]{Stilpa,Mois}. As stated earlier, turbulence can compress gas over scales below the Jeans length and cascade downward to produce very small structure in the ISM, leading to star formation. Turbulence can also quench the star formation activities since it can disrupt gaseous structures faster than they gravitationally collapse. On the other hand, theories also suggest that star formation feedback can drive supersonic turbulence through expanding \ion{H}{2} regions, although this is not well understood \citep{os11}.

In regions of neutral hydrogen, the timescale involved in the progression from \ion{H}{1} to molecular clouds to star formation can be long. This is estimated to be about $30$ to $40$~Myr for dwarf galaxies \citep{Stilpb}, making the connection of turbulence and star formation less direct. Nevertheless, in this paper we present our study of the higher order statistics (HOS), skewness and kurtosis, of the \ion{H}{1} maps for a sample of spiral galaxies to examine whether a correlation between turbulent gas motions and star formation exists.

Our first application is to compute the moments of the whole galaxy and compare with the integrated SFR, as shown in Figure~\ref{kvssfr}. This comparison does not suggest any correlation among the galaxies as a whole. Next we compare kurtosis to FUV surface photometry as a function of radius, shown in Figure~\ref{kvsfuv}. These plots also do not appear to draw a firm correlation in our sample. The FUV surface photometry in most galaxies shows a downward or nearly flat trend with radius. The kurtosis, however, shows a peak at different radius for some galaxies, and in others it follows a loosely upward trend. We conclude that no general correlation between the FUV surface photometry and kurtosis as a function of radius can be found.
However, we note that when performing the annulus analysis, averaging azimuthally will cause the variations in \ion{H}{1} distribution within the annulus to be lost. This could potentially wash out the correlations that we are looking for, if the scale of star formation is smaller than our annuli width.

Finally, we consider HOS and FUV emission divided in a grid map, as shown in Figure~\ref{gridmaps}. Each kurtosis map is almost uniform across the galaxy, while maps of skewness show a greater variation with higher values around the edges. Skewness appears to trace the arms in some galaxies, but the correlation is not consistent across the sample. We find that neither kurtosis nor skewness maps show a strong correlation to the FUV maps in this application.

Even though there is no apparent correlation between HOS and FUV in our three different applications, the interpretation may not be so straightforward, and we note that in-depth analysis may be required to make a more direct comparison in the future. Models of turbulent-based star formation suggest that star formation rate scales with the gas surface density (such as the Kennitcut-Schmidt law), and that the fraction of gas in collapsing structures, which infers the star formation rate, is only a very weak function of the Mach number and other properties of the turbulent flow \citep[e.g.~][]{elm02,krumholz,Ren}. Therefore future analysis may consider a comparison based on fixed gas surface density. Nevertheless, we find that the analysis of HOS and sonic Mach number is indeed applicable to spiral galaxies. Specifically, our analysis of HOS in the whole image and in a grid suggest that gas motions across each galaxy are supersonic both globally and on a smaller scale.

\subsection{Potential Resolution Effects} \label{resolution}

A significant factor that could contribute to the lack of correlation between HOS and FUV in our applications is the spatial resolution of our analysis. When performing our annulus analysis, in order to ensure enough signal per annulus, we choose the increment in radius to be $20$~pixels for most galaxies (30~pixels for NGC~2403), which corresponds to a scale of $\sim700$~pc at an average distance of $5$~Mpc for our galaxies. It is possible that this may have been too low of a spatial resolution to resolve the variations in HOS and FUV we were looking for. 

For example, in Figure~\ref{kvsfuv} the FUV magnitudes show an increase at $\sim5$~kpc for NGC~3031. A slight increase in FUV can also be seen in galaxies NGC~6949 and IC~2574 between $5$ and $10$~kpc. When comparing to the FUV images, this increase can be associated with large changes in FUV brightness over a large distance. For example, in NGC~3031 this increase of FUV magnitudes at $\sim5$~kpc coincides with passing from the relatively empty inner ring to the beginnings of the spiral arm pattern. In NGC~6946 this FUV increase at $\sim6$~kpc occurs at a similar change in the galaxy. There are large sections of empty pixel within the galaxy, so the increase in FUV emission as a function of radius can be explained by passage from a relatively FUV empty annulus to the next annulus where there is more FUV emission. 

These FUV variations are easily detected and can be explained when examining kpc scales across a galaxy. However any variations on a smaller scale are very likely not visible in our analysis. Turbulence, and therefore the HOS, might vary on a scale smaller than the spatial resolution of our analysis. If our annulus analysis at the scale of $20$~pixels cannot resolve the HOS and FUV variations, the same effect would appear in our grid analysis, which compares $32\times32$~square pixels in the galaxies. This may indicate an upper limit of $\sim700$~pc for the scale of observed turbulent motions.

Both theoretical and observational studies suggest that the nature of turbulence and its driven mechanisms are different below and above the scale of line-of-sight disk thickness, which are typically $\sim100$~pc in nearby disk galaxies \citep{elm03,block,Bourn,combes12}. The energy injection in the ISM on scales $<100$~pc mainly comes from gravitational processes, such as gravitational instabilities and inward mass accretion \citep{Bourn,elm10,kles10}, with regulations from stellar feedback, and the behavior of turbulence in these scales is three dimensional. Therefore in our samples the turbulence presumably is two dimensional driven by disk self-gravity or global disk rotation, and mixed with density-wave and bar-driven streaming motions \citep{block,Bourn}.

\subsection{Gas Motion in Interacting Galaxies}

Studies of interacting galaxies, both theoretically and observationally, suggest that shocks triggered during the process can induce star formation in these galaxies \citep[e.g.~][and references within]{elme95,degrijs03,barnes04,chien07,chien10}. Two of the galaxies in our sample, NGC~3031 (or M81) and NGC~5194, are in the process of interacting with another galaxy. In particular, NGC~3031, M82, and NGC~3077 is a system of three interacting galaxies. The analyses of HOS and FUV for these two galaxies do not appear to be different from other galaxies in our sample. However, the cause of turbulence in the gas of these galaxies may be the interaction itself, which may possibly result in a different spatial distribution from those in single spiral galaxies. This is an open question we are investigating and a similar study on more interacting galaxies is underway.

\subsection{Summary and Future Work}
1. Our analysis does not indicate any significant correlation between turbulence and star formation in spiral galaxies. Distributions of statistical moments across a galaxy do not trace any star-forming regions. However, we note that our analysis, and thus the possible interpretations, can be improved through a more in-depth comparisons based on fixed gas surface density among the galaxies.

2. We find that the analysis of HOS and the sonic Mach number in \citet{Burk} is applicable to spiral galaxies as well, and that the gas motions in each galaxy are largely in the supersonic regime. These gas compressions may be due to turbulence, but also likely due to local shearing or shocks. In the future we would like to map the Mach number across the galaxies, in order to see its distribution in the ISM, and analyze whether there is a relation of the cold neutral medium, the warm ionized medium and star-forming regions \citep{yl96,y03,deBlok}. 

3.  Much of our analysis may have been resolution limited--- variations in HOS that would indicate the correlations of turbulence and star formation may only be detectable at smaller spatial scales. On the other hand, this also indicates that our analysis possibly places an upper limit of $\sim700$~pc on the scale of turbulent motions. Presumably on such large scales ($>100$~pc), turbulence is driven by disk gravitational instabilities and mixed with density-wave or bar-driven streaming motions. Therefore in our samples, we may be observing large-scale turbulence that is less associated with stellar activities. Future work may include more similar analysis at higher spatial resolutions, if possible, without compromising the signal-to-noise ratio. One way to increase the resolution of the grid analysis is to overlap each kernel with the previous, without reducing its size, rather than creating discrete kernels.

4. The same analysis for a larger sample of dwarf galaxies from the LITTLE THINGS\footnote{Local Irregulars That Trace Luminosity Extremes, The \ion{H}{1} Nearby Galaxy Survey: an \ion{H}{1} survey of 21 nearby dwarf irregular and Blue Compact Dwarf galaxies taken using the VLA} \citep{Hunt} survey is underway. We will compare the results, in a separate paper, to see what further correlations can be drawn between dwarf galaxies and spiral galaxies. A similar study of interaction-triggered shocks, turbulence, and star formation in interacting galaxies is also in progress.

\acknowledgments
The authors gratefully acknowledge the editor and the helpful comments and encouragement from the referee. We also like to acknowledge the support from the National Science Foundation through grant number AST-1461200, awarded to Northern Arizona University, Department of Physics and Astronomy. DAH is grateful to John and Meg~Menke for funding for page charges. E.~Maier and LHC would like to thank Dr.~Kathy~Eastwood for her help and support.


\end{document}